\documentclass[11pt,preprint,authoryear,a4paper]{elsarticle}

\usepackage{journalnames}
\usepackage{natbib}
\usepackage{amssymb}
\usepackage{graphicx}
\usepackage{epsfig}
\usepackage{boxedminipage}

\begin{document}

\begin{frontmatter}

\title{N-body Integrators with Individual Time Steps from Hierarchical Splitting}

\author[lei]{Federico I. Pelupessy}
\ead{pelupes@strw.leidenuniv.nl}
\author[ams,cam]{J\"urgen J\"anes}
\author[lei]{Simon Portegies Zwart}

\address[lei]{ Leiden Observatory, Leiden University, PO Box 9513, 
2300 RA, Leiden, The Netherlands}
\address[ams]{ Faculty of Science, University of Amsterdam, PO Box 94216,
1090GE, Amsterdam, The Netherlands}
\address[cam]{ Cambridge Computational Biology Institute, Department of Applied
Mathematics and Theoretical Physics, Centre for Mathematical Sciences,
Wilberforce Road, Cambridge CB3 0WA, United Kingdom}


\begin{abstract}

We review the implementation of individual particle time-stepping for N-body
dynamics. We present a class of integrators derived from second order
Hamiltonian splitting. In contrast to the usual implementation of individual
time-stepping, these integrators are momentum conserving and show excellent
energy conservation in conjunction with a symmetrized time step 
criterion. We use  an explicit but approximate formula for the time 
symmetrization that is compatible with the use of individual time steps.  
No iterative scheme is necessary. We implement these ideas in the 
HUAYNO\footnote{available online at www.amusecode.org} code and present 
tests of the integrators  and show that the presented integration schemes shows 
good energy conservation, with little or no systematic drift, while conserving
momentum and angular momentum to machine precision for long term
integrations.

\end{abstract}


\end{frontmatter}

\section{Introduction\label{sec:intro}}

Astrophysical N-body problems often show a large dynamic range of 
timescales within their component systems. Instead of a fixed or varying
global time step most current astrophysical codes update the force and
advance particles with a time step that is determined for each particle
separately \citep[for a recent review see][]{Dehnen2011}. Such individual particle
time step schemes (IPTS) allow efficient simulations by concentrating the
computational resources on the parts of the system that experience the
strongest evolution. The most popular IPTS is a block time step scheme where
the particles are organized in a power of two hierarchy of time step bins
\citep{McMillan1986, Hernquist1989, Makino1991}. The bin in which a particle
resides determines the frequency of force evaluations and
each successively higher bin has a factor 2 smaller interval between updates. 
This block time step scheme has been adopted in a
wide range of codes, in for example  codes for galactic simulations
\citep{Dubinski1996, Magorrian2007}, cosmological simulations
\citep{Stadel2001},  Smooth Particle Hydrodynamics  codes
\citep{Springel2005, Wadsley2004} and also modern Hermite integrators for
collisional stellar systems  \citep{Makino1991, Aarseth1999,
PortegiesZwart2001, Harfst2008, Konstantinidis2010}. The popularity of  this
scheme stems from the fact that it allows for individual tailored  time steps
while still grouping particles with similar time steps together - which means
that the cost of synchronizing the rest of the system is shared by all the
particles in a given bin and parallelization of the force calculation is
possible.

A notable exception where the simple block time step scheme is not used 
is when long term integration requires conservation of the integrals of 
motion to high precision. A prototype of a problem where this is required 
is the integration of planetary systems over billions of years. In this 
case care must be taken because IPTS typically do not conserve 
momentum and show drifts of the total energy. For these problems symplectic 
integrators based on Hamiltonian splitting \citep{Wisdom1991} are used,
using the fact that the dynamics is dominated by a central object. For close
encounters between objects orbiting the central body, additional splitting
methods have been developed involving the interaction terms 
between orbiters \citep{Chambers1999, Duncan1998, Moore2011}. 

Here we will derive a class of IPTS integrators applying these ideas  for
general N-body problems by splitting the N-body  Hamiltonian starting from
an initial pivot time step. We subdivide particles in two groups, according
to whether their time step is shorter or longer than the pivot. The time
evolution of the slow particles  is evaluated using the current time step,
where the terms involving  particles with a shorter time step are grouped
into a new Hamiltonian to which we apply the same procedure twice with a
halved pivot time step.  This procedure is repeated recursively until the
time evolution of all particles is evaluated.  In this way different
integrators can be derived (depending on the details of the split) that
retain a close resemblance to the block time step scheme, but which
will conserve integrals of motion, in particular the total (angular) momentum, 
to machine precision. We will show that, when used in conjunction with an 
approximately symmetric time step criterion, they show good energy conservation 
and do not show energy drift. 
 In section~\ref{sec_method} we present our derivation and our
symmetrization scheme . In section~\ref{sec_tests} we present  various tests
of these integrators, while we summarize and discuss the  results in section
\ref{sec_disc}.

\section{Method}
\label{sec_method}

\subsection{Derivation}
\label{sec_der}

The problem at hand is to evolve a set of particles X for a time $dt$ under
the dynamics generated by the Hamiltonian:
\begin{equation}
    H = \sum_{i \in X} \frac{p_i^2}{2 m_i} - 
        \sum_{i,j \in X,\ i<j} \frac{G m_i m_j}{| r_i - r_j|}
\end{equation}
The Hamiltonian consists of momentum terms
\begin{equation}    
    P_X \equiv \sum_{i \in X} \frac{p_i^2}{2 m_i} 
\end{equation}  
and potential terms 
\begin{equation}
    V_{XX} \equiv - \sum_{i,j \in X,\ i<j} \frac{G m_i m_j}{| r_i - r_j|}
\end{equation}
The evolution of the state of the system is given by the flow 
operator $\exp( dt \mathbb{H})$ where $\mathbb{H}$ is the Hamiltonian
vector field corresponding to $H$. In cases the Hamiltonian can be 
split in two parts $H=H_A + H_B$, where $\exp{dt \mathbb{H}_A}$ and 
$\exp{dt \mathbb{H}_B}$ can be calculated, the flow of $H$ can be 
approximated by composing $\exp{dt \mathbb{H}_A}$ and $\exp{dt \mathbb{H}_B}$,
e.g.
\begin{equation}  
    \exp( dt \mathbb{H} ) \approx \exp(dt/2 \mathbb{H}_B) \exp(dt
    \mathbb{H}_A) \exp(dt/2 \mathbb{H}_B)
\end{equation}
in this case the approximation holds to second order in $dt$
\citep{Hairer2006}. An example of this is the division in $H_A=V_{XX}$ and 
$H_B=P_X$ which generates the familiar second order Drift-Kick-Drift 
(DKD) verlet integrator.

The derivation of approximate solvers by the division of the Hamiltonian 
into simpler, solveable, subsystems is a general method not limited to the
above division in momentum and potential terms. For example, efficient
planetary integrators may be constructed by splitting the Hamiltonian in a part
that describes the Keplerian motion around the central star and a part that 
describes the perturbation of the planetary bodies on each other~\citep{Wisdom1991}. 
Additionally, splitting can be used as a strategy for focusing the computational 
effort on the parts of the system that are evolving most strongly, either by 
decomposing the potential in two \citep[][for a split into long range and short 
range forces in cosmological simulations]{Springel2005} or more components 
\citep[][employing a series of succesive shells]{Duncan1998}, or by grouping 
particles according to their dynamical evolution timescale 
\citep[e.g.][]{Fujii2007, Saha1994}. Here we derive multiple time step methods 
by choosing the division of the Hamiltonian \emph{adaptively and recursively based 
on the current time step assigned to the particles} 
such that one system S contains the interaction terms
of all the particles with a time step larger than $dt$, and the other system F
contains all the interaction terms of the particles with a time step smaller 
than $dt$. This method of subdividing the system according to the timestep 
is very close to splitting in ~\cite{Saha1994} (the ``PASS'' method we present 
below employs the same split as their method),  apart from the fact that we allow 
the partitioning of the system to change continuously. This last property means that
we have to be careful choosing the timesteps (discussion of which we defer to 
section \ref{sec_symtime}).

For example, we may choose the split
\begin{eqnarray}  
\label{eq_hold}
  H_S & = & P_S + V_{SS} + V_{SF} \nonumber \\
  H_F & = & P_F + V_{FF} 
\end{eqnarray}  
(Note that indeed $H=H_S+H_F$) 
and approximate  
\begin{equation}  
\label{eq_SF2nd}
    \exp( dt \mathbb{H} ) \approx \exp(dt/2 \mathbb{H}_F) \exp(dt \mathbb{H}_S) \exp(dt/2 \mathbb{H}_F)
\end{equation}  
The evolution operator $\exp{dt \mathbb{H}_S}$ is then approximated by 
the DKD (or any other second order scheme). This consists of drifts of the
particles in S (from $P_S$) and kicks on the particles in S \emph{and} F 
(from $V_{SS}+V_{SF}$). These kicks consists of updates of the
particle positions where every $-G m_i m_j / | r_i - r_j |$ term in the
potential part generates kicks 
\begin{eqnarray}
dp_i & = & dt \frac{G m_i m_j}{|r_i - r_j|^3} (r_i -r_j) \nonumber \\
dp_j &= & dt \frac{G m_i m_j}{|r_i - r_j|^3} (r_j -r_i)
\end{eqnarray}
For the two $\exp{dt/2 \mathbb{H}_F}$ operators of the F system we apply 
the same procedure as for the $\exp{dt \mathbb{H}}$ operator (with halved 
time step). The splitting can thus be applied recursively to the evolution 
operators with time step $dt/2^k$. The recursion ends when a time step
sufficiently small is reached such that no particles end up in the F system
(of that level).  Note that we made a choice 
in subdividing  the system: Eq.~\ref{eq_hold} is certainly not the only possible split. 
For example, the potential cross terms
$V_{SF}$ could also be put in $H_F$, resulting in a different method.
The difference between the two is that in the former case
(eq.~\ref{eq_hold}) the interactions between particles in $S$ and $F$ systems 
are done on the slowest of the time steps of the i and j particle pair.  
In the latter case the interactions propagate to the fast
system, and will in the end be updated on the minimum of the time steps 
of each $(i,j)$  particle pair, thus the split in eq.~\ref{eq_hold} can 
be expected to be computationally faster. Note that the integration scheme 
that results is similar to conventional individual block time step schemes,
where the time step is subdivided in powers of two. However, 
existing direct N-body codes with block time step schemes 
\citep[e.g.][and modern variants]{McMillan1986, Makino1991} put 
particles in a hierarchy of time step bins to dictate the frequency in 
which the \emph{total} forces for the particles
in that bin are updated. A particle in a fast time step bin is evolved using 
a force calculated from all other particles - the particles that need less 
updates (higher in the time step hierarchy) are extrapolated to the time of
the fast particle if necessary. In the scheme presented here 
the total force is \emph{never} (or only incidentally) calculated: all the 
velocity changes are effected by momentum kicks (which need only originate 
from a subset of the particles) - and always pairwise: every 
$i \leftarrow j$ interaction kick is accompanied by its 
corresponding $i \rightarrow j$ interaction kick. The scheme is therefore
manifestly momentum conserving. The extrapolating schemes, including their 
Hermite generalizations, are not momentum conserving.

\begin{figure}
\centering
\epsfig{file=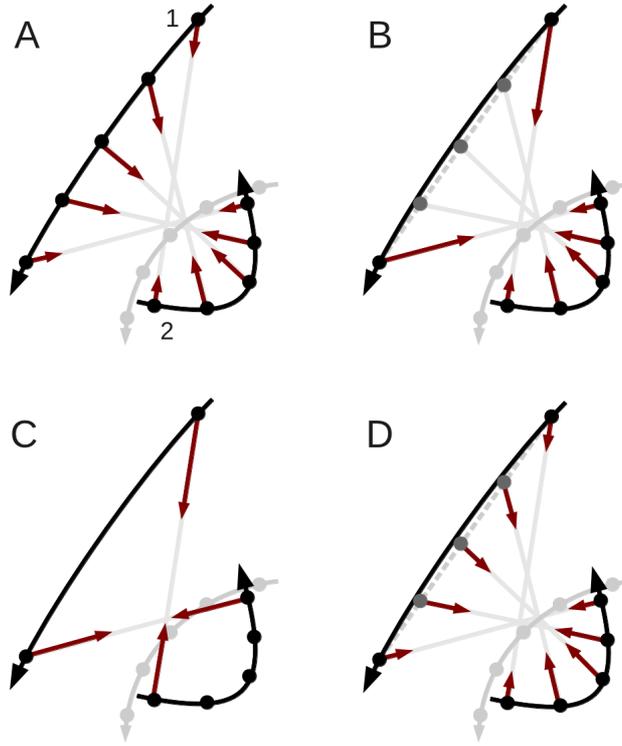, width=.75\textwidth}
\caption{
 Schematic overview of the different integrators. Panel A shows the
 interactions that are calculated between two particles (1 and 2) by the
 SHARED integrator for a fiducial trajectory of these two particles, where
 particle 1 has a time step $4\times$ larger than particle 2, which is also
 globally in the smallest time step bin (with the rest of the system  is
 represented by the greyed out trajectory). All the interactions between the
 particle are calculated on the smallest time step. Panel B shows the case of
 the conventional block time step scheme (BLOCK integrator). In this case the
 total force on each particle is calculated determined by the individual
 time step of that particle. Panel C shows the HOLD integrator 
 (Eq.~\ref{eq_hold}). In this case particle 2 only feels the interaction  of
 particle 1 with a frequency determined by the time step of particle 1. In
 between the particle may be kicked by interactions with other particles. 
 Panel D shows the PASS integrator (Eq.~\ref{eq_pass2}). For this
 integrator the mutual interactions  between 1 and 2 are calculated on a
 time step determined by the fast particle 2 (note that in this figure the
 same number of interactions is  calculated for the SHARED and PASS
 integrators - in general this is not  the case). The difference with panel
 B is the fact  that each kick of the velocity of particle 2 is matched by
 the corresponding kick on particle 1. 
 }
\label{fig_interface}
\end{figure}

By putting the $V_{SF}$ in the slow system in Eq.~\ref{eq_hold} we made the choice of 
calculating the interactions between `slow' and `fast' system on the
time step of the slow particles - hence we refer to this integrator as the HOLD 
integrator (the $V_{SF}$ are `held' on the slow time step). Alternatively,
when we put the $V_{SF}$ terms in the fast system:
\begin{eqnarray}  
\label{eq_pass1}
  H_S & = & P_S + V_{SS} + V_{SE} \nonumber \\
  H_F & = & P_F + V_{FF} + V_{FS} + V_{FE}
\end{eqnarray}  
Note that we have included the $V_{SE}$ and $V_{FE}$, these denote any
interactions that are inherited from the higher levels. As we will see below
this integrator does not conserve the position of the center of mass, the
corresponding integrator that does is  
\begin{eqnarray}  
\label{eq_pass2}
  H_S & = & V_{SS} + V_{SE} \nonumber \\
  H_F & = & P_S + P_F + V_{FF} + V_{FS} + V_{FE}
\end{eqnarray}  
We refer to this scheme as PASS (and to eq.~\ref{eq_pass1} as PASS1).

We will compare the HOLD and PASS integrators with the conventional 
integrator without individual time steps but with shared adaptive time steps
(SHARED) and the conventional block time step integrator which extrapolates
particle positions (BLOCK). A cartoon representation and overview of the
differences  between the different integrators is given in
Figure~\ref{fig_interface}.

\subsection{Symmetric time-stepping}
\label{sec_symtime}

The above integrators are symplectic only as long as the particles do not 
change their timestep. For general N-body simulation codes, where particles 
experience very different dynamical timescales during their evolution, we 
need to relax this condition and let the particles move between timestep sets.
Using adaptive timesteps dependent on the phase space coordinates will in 
general destroy the symplecticity (and hence the conservation properties) 
of the integrator \citep{Skeel1992,Preto1999, Mikkola1999}, although with some restrictions 
it is possible to construct adaptive timestep symplectic integrators by 
considering the Hamiltonian in an extended phase space \citep{Preto1999, Mikkola1999} or 
using a variational approach \citep{Farr2007}. For the integrators we present 
here it is however possible to recover long term energy conservation  
by ensuring time reversibility of the integrator. This can be done by using a 
time-symmetrized time step criterion for the particles \citep{Hut1995, Preto1999}.

To be specific, we start by reviewing in more detail why the numerical 
integration of a particle trajectory loses its time symmetry (and hence 
its energy  conserving properties) when a variable time step
is adopted. Given a particle trajectory $(r(t),v(t))$ one tries numerically
to integrate this using a time step function $\tau(t)$ (which is a function
of  time through its dependence on the state variables $r$ and $v$). If
we advance a state $r(t)\rightarrow r(t+\tau_1),v(t) \rightarrow
v(t+\tau_1)$ we reach a new state with a new time step $\tau_2$. If we
would run the integration backwards in time, in general $\tau_1 \ne \tau_2$, so 
we would not arrive at the original state, and due to the numerical errors 
we would not follow the same numerical approximation to the trajectory. 
This situation would not arise if the
time step function had the property that $\tau(t+\tau(t))=\tau(t)$. 
Note that $\tau(t)$ refers to a time step in positive direction 
while $\tau(t+\tau(t))$ refers to a time step in the negative direction, 
so the requirement should be read as $\tau^-(t+\tau(t))=\tau^+(t)$. 
In fact we can introduce such a symmetric time step function 
$\tau_{\rm sym}$:
\begin{eqnarray} 
\label{eq_symdt} 
   \tau^-_{\rm sym}(t) & = & \tau(t)/2 + \tau(t-\tau^-_{\rm sym})/2 \nonumber\\    
   \tau^+_{\rm sym}(t) & = & \tau(t)/2 + \tau(t+\tau^+_{\rm sym})/2 
\end{eqnarray}  
this becomes a workable ansatz if we approximate 
\begin{equation}  
   \tau^+_{\rm sym}(t) \approx \tau(t) + \frac{1}{2} \frac{d\tau}{dt} \tau^+_{\rm sym}
\end{equation}  
so we can write:
\begin{equation}
\label{eq_symdt2}  
\tau_{\rm sym}= \frac{\tau(t)}{(1-\frac{1}{2} \frac{d\tau}{dt})} 
\end{equation}
which is the approximation we will use below. 
For a time step proportional to the inter-particle free-fall times
\begin{equation}  
  \label{eq_ff}
  \tau_{ij}=\eta_1 \sqrt{\frac{r_{ij}}{a_{ij}}}=\eta_1 \sqrt{\frac{r^3_{ij}}{\mu_{ij}}}
\end{equation}  
with $\mu_{ij}=m_i+m_j$, the derivative is  
\begin{equation}  
  \label{eq_ffderiv} 
  \frac{d\tau_{ij}}{dt}=\frac{3 \overline{v}_{ij} \cdot \overline{r}_{ij}}{2 r_{ij}^2} \tau_{ij} 
\end{equation}  
hence the symmetrized version of the free-fall time step criterion becomes
\begin{equation}  
  \label{eq_symts}
  \tau_{i}= \min_j \left( \frac{\tau_{ij}}{(1-\frac{1}{2} \frac{d\tau_{ij}}{dt})} \right) 
\end{equation} 
A time step proportional to the interparticle flyby times
\begin{equation}  
  \label{eq_fb}
  \tau_{ij}= \eta_2 \frac{r_{ij}}{v_{ij}}
\end{equation}  
with 
\begin{equation}  
  \label{eq_fbderiv}
  \frac{d\tau_{ij}}{dt}=\frac{\overline{v}_{ij} \cdot \overline{r}_{ij}}{r_{ij}^2} \tau_{ij}
  (1+\frac{\mu_{ij}}{v_{ij}^2 r_{ij}}) 
\end{equation}  
can be similarly made symmetric (note that eq.~\ref{eq_fbderiv}
simplifies the expression by only considering the $i$ and $j$ particle for
the time derivative of $v_{ij}$). We will take the particle time step 
to be equal to the minimum of the symmetrized free-fall (defined by
eqs.~\ref{eq_ff}, \ref{eq_ffderiv}, and \ref{eq_symts}) and 
symmetrized flyby time step (eq. \ref{eq_fb}, \ref{eq_fbderiv} and
\ref{eq_symts}). 

There are various other ways in the literature to symmetrize the time step.
\cite{Hut1995} advocated an iterative procedure to symmetrize
Eq. \ref{eq_symdt}. In this case an trial step is taken, after which the 
average of the new and old time step is taken. This can be repeated to 
converge to the solution of Eq. \ref{eq_symdt}. This has the drawback that 
the iteration is an expensive operation which needs multiple force 
evaluations for the system. Alternatively, a time step determined by
\begin{equation}
\tau_{\rm old} \tau_{\rm new} = \tau(t)^2
\end{equation}
provides an explicit symmetrized time step \citep{Dehnen2011}. Both of these do
not work when used with a block time step scheme \citep{Makino2006,
Dehnen2011}. For the former iterative solution this is because of a
``flip-flop'' problem where convergence  is not reached because of the
restriction of the time step to powers of two of a base time step. A solution
in case is available \citep{Makino2006}, but entails saving the whole
system state over a time interval and multiple iterations.  

\begin{figure}
\centering
\epsfig{file=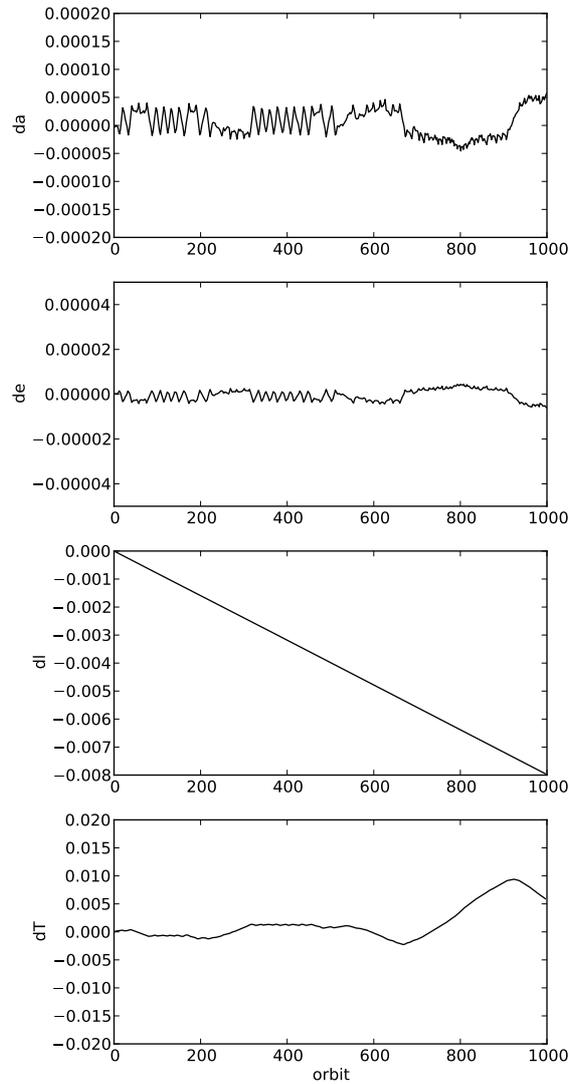, width=.6\textwidth}
\caption{Orbital elements for the variable time step leapfrog integrator with
symmetrized time step criterion (eqs.~\ref{eq_ff}, \ref{eq_ffderiv}, and
\ref{eq_symts}). Plotted are the changes in semi-major axis a, eccentricity e,
 longitude of apo-center l and the time of apocenter passage as a function of
 orbit number (sampled at apocenter). The time steps are varied in the same
 way as in the blocktime step schemes. }
\label{fig_kep1}
\end{figure}

To test whether the expressions given here (which due to the truncation after
first order in $\tau$ in Eq.~\ref{eq_symdt2} are only approximately time-symmetric) 
yield a conservative variable
time step method when using a power of two hierarchy of time steps, we
integrate a Kepler  orbit with eccentricity $e=0.9$ for 1000 orbits and a
$\eta=0.01$  \citep[this is the same test as fig. 1 and 2 of][]{Hut1995}. 
The test uses the symmetrized free-fall time step (eq.~\ref{eq_ff}). The 
time steps are chosen to be block time steps as in the integrators presented 
in section ~\ref{sec_der} (for the two body problem the integrators of
section~\ref{sec_der} and the usual block time-stepping are equivalent). 
The orbital elements for the orbit are plotted in Fig.
\ref{fig_kep1} . As can be seen there is no systematic drift in energy (the
orbit does precess with a constant speed). The performance of the integrator
is similar to the iterative scheme of \citep{Hut1995}, and does not show the
systematic drift in $a$ and $e$ of a non-symmetrized integrator.

\begin{figure}
\centering
\epsfig{file=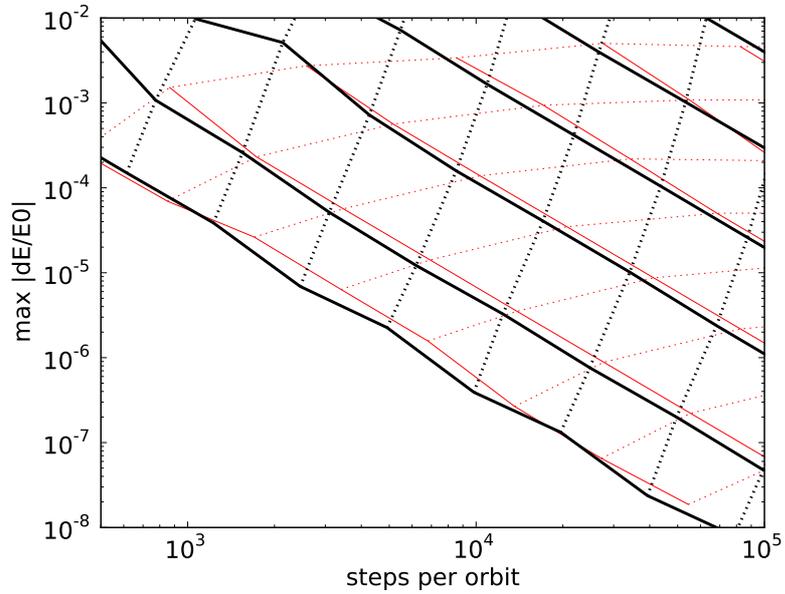, width=.89\textwidth}
\caption{Maximum relative energy error vs number of steps per orbit. For
this a Kepler orbit is integrated for 10000 orbits. The lines in black use
the standard symmetrized time step criterion (eqs.~\ref{eq_ff}, \ref{eq_ffderiv}, and
\ref{eq_symts}) for different values of the eccentricity (from bottom to top
$e = 0.6,0.9,0.99,0.999,0.9999,0.99999 $), dotted lines connect runs with 
the same $\eta$. Thin red lines show the same for runs with the error
control time step Eq.~\ref{eq_ec}.}
\label{fig_kep2}
\end{figure}

In Figure~\ref{fig_kep2} we show the resulting (maximum) energy error for 
a set of $10^4$ orbit integrations for different values of the eccentricity
and time step parameter $\eta$, plotted vs the number of steps per orbit. 
Although the integration scheme is in general robust, this figure shows that
for a given $\eta$ it tends to keep the number of steps per orbit constant.
This could be a disadvantage if such a time step is used 
in a run with multiple bodies where encounters with different $e$ can occur.
This can be remedied by choosing an error control time step $\tau_{ec}$
\begin{equation}
\label{eq_ec}
\tau_{ec}=\tau (\frac{\tau}{\tau_0})^\gamma
\end{equation}
with $\gamma$ a small exponent such that the time steps become smaller as
$\tau$ decreases. Such an expression is also easily symmetrized. 
Figure~\ref{fig_kep2} also shows the results for this time-stepping scheme 
(with $\gamma=1/3$). As can be seen, in this case runs with constant $\eta$
show a much less marked increase in error, showing an almost flat profile
for the orbits with extreme $e$ (albeit at the price of a rapidly increasing
number of steps per orbit) For the other tests below we do not use
the $\tau_{ec}$ as these tests are not especially sensitive to the high 
eccentricity behavior.

\subsection{Implementation}

\begin{figure*}
\caption{ Pseudo code of the HOLD integrator.}

\begin{boxedminipage}{10cm}
{
\small
\begin{verbatim}
function evolve_hold(system, dt)
    calculate_timestep(system)
    fast, slow=sort_in_fast_and_slow(system,dt)
    evolve_hold(fast, dt/2)
    drift(slow,dt/2)
    kick(slow, fast, dt)
    kick(fast, slow, dt)
    drift(slow,dt/2)
    evolve_hold(fast, dt/2)
\end{verbatim}
}
\end{boxedminipage}
\label{fig_C}
\end{figure*}

The integrators are implemented in the HUAYNO (for  Hierarchically split-Up
AstrophYsical N-body sOlver) N-body code, released as part of the
Astrophysical Multi-purpose Software Envrionment
\cite[AMUSE,][]{PortegiesZwart2009}\footnote{www.amusecode.org}. 
HUAYNO is written in C as testbed for the
N-body integrators described here. There is a very close correspondence of 
procedures in HUAYNO and the mathematical operators for the algorithm: 
evolve, drift and kick operators have their counterpart in the code (this is
illustrated in the pseudocode for the HOLD integrator given in
fig.~\ref{fig_C}). 
The particles themselves are stored in an array
which is reshuffled to match the partitioning required by the integrator 
\citep[this partitioning is very similar to the block time step
implementation of][]{McMillan1993}. The subsystems are then 
contiguous memory blocks conveniently referred to by a pointer and size. 
This has the added benefit that the particles which end up accessed the most
end up close to each other in memory. The shuffling itself is an $O(N)$ 
operation with negligible impact on the performance. 

The code implements an optional Plummer softening parameter $\epsilon$ for the 
gravitational interactions. It allows to set separately $\eta_1$ and $\eta_2$,
for the time step parameter of the free-fall (eq.~\ref{eq_ff}) and flyby
(eq.~\ref{eq_fb}) time step criteria. If not explicitly mentioned they are 
taken to be the same (and referred to by $\eta$) in the tests below. 
Unless explicitly given the tests below use unsoftened gravity, $\epsilon=0$.

\section{Tests} 
\label{sec_tests}

\begin{figure}
\centering
\epsfig{file=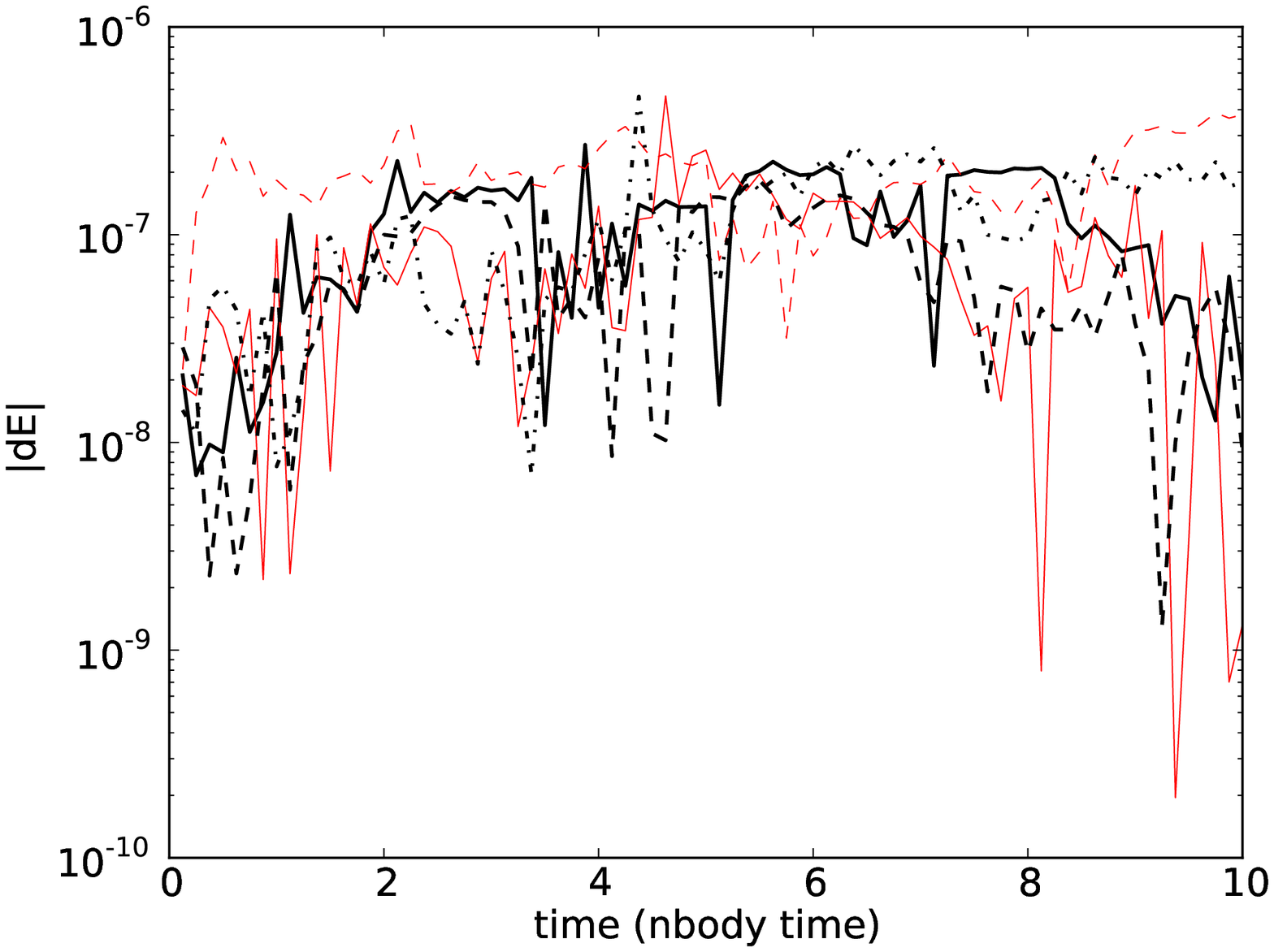, width=.49\textwidth}
\epsfig{file=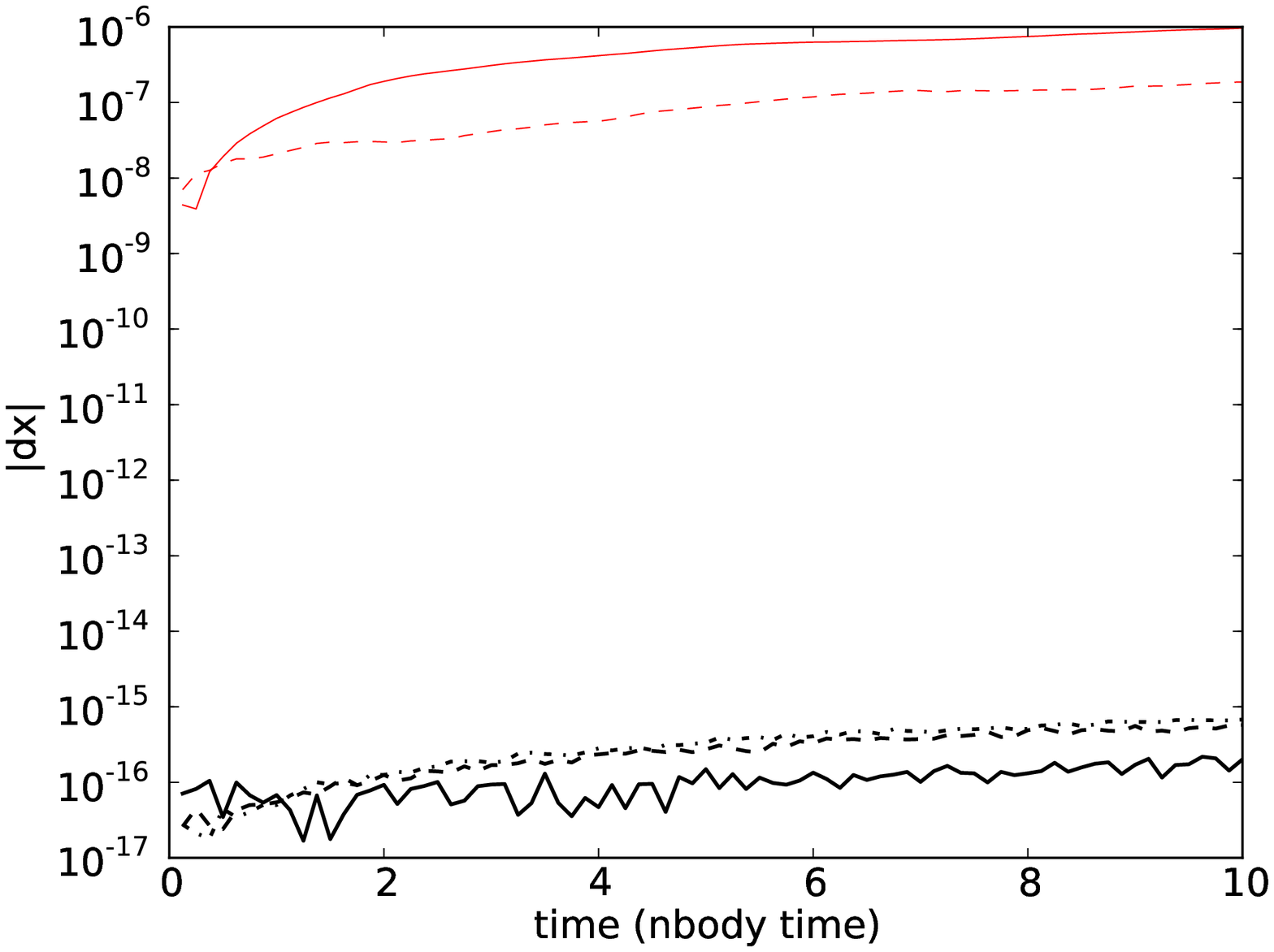, width=.49\textwidth}
\epsfig{file=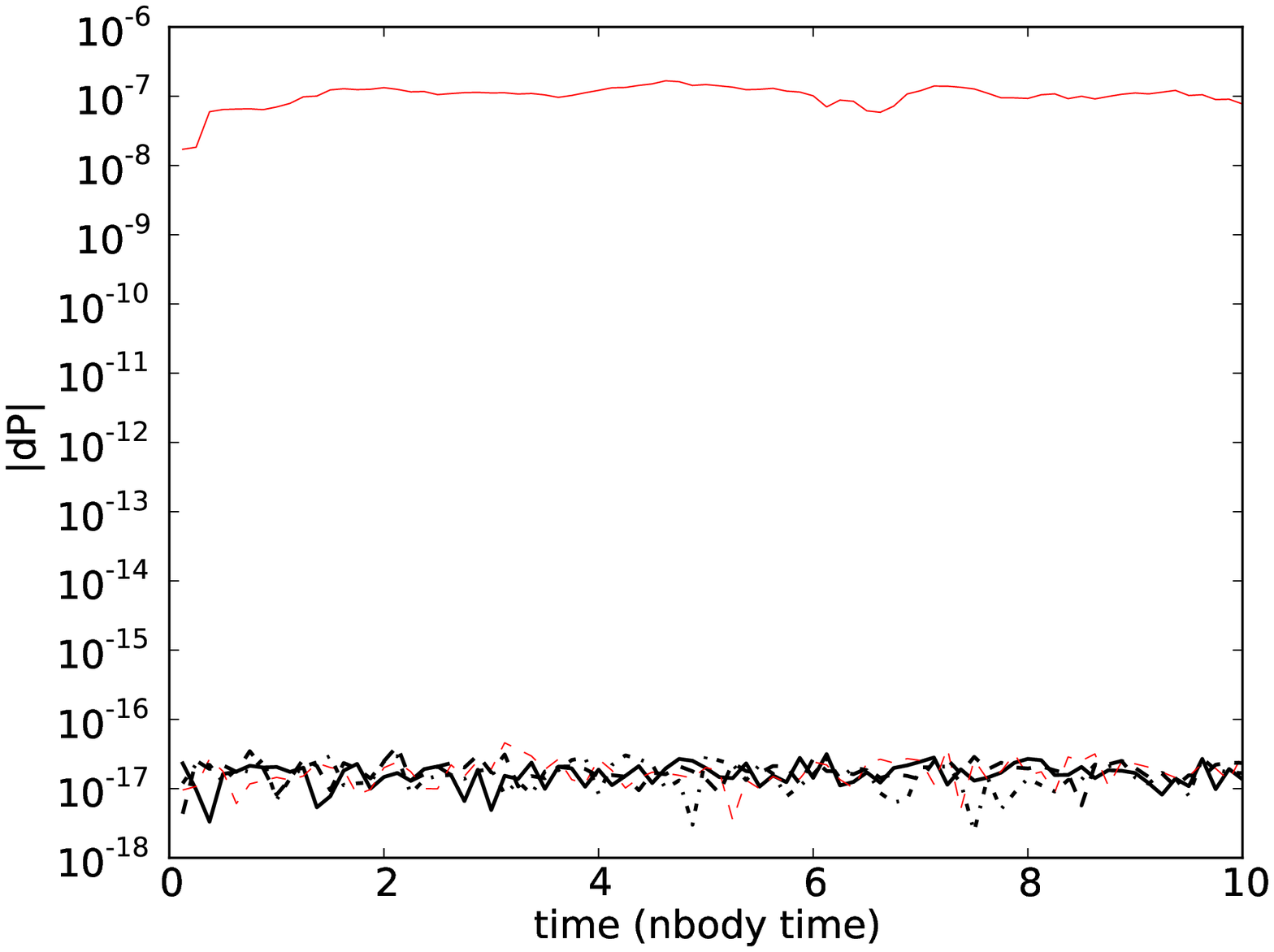, width=.49\textwidth}
\epsfig{file=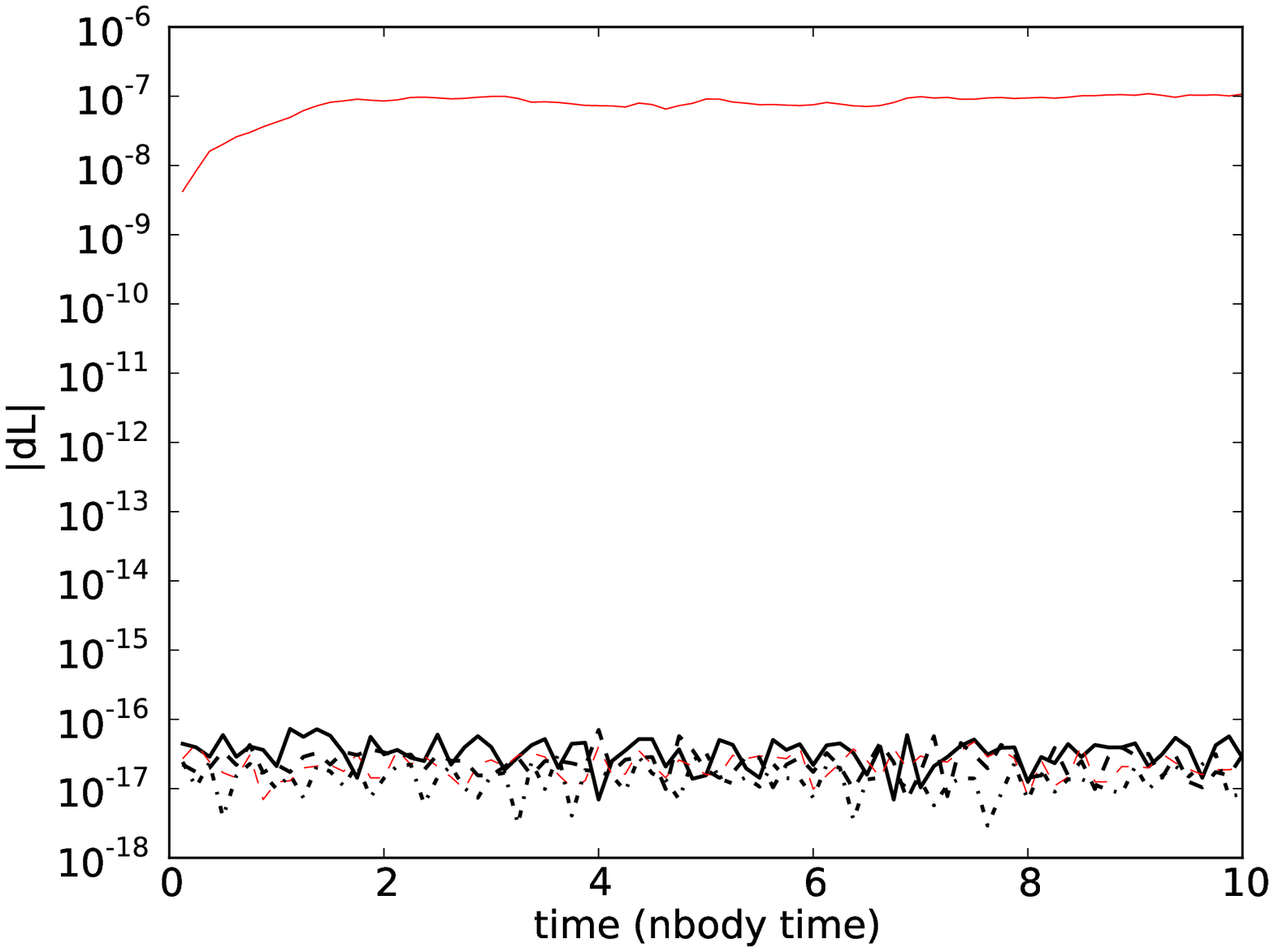, width=.49\textwidth}
\caption{Error in conserved quantities for $N=100$ plummer sphere runs.
Shown is the energy error (upper left panel), drift in the center of 
mass position (upper right), linear momentum (lower left) and angular 
momentum (lower right) for different integrators: SHARED (thick black line),
BLOCK (thin red line), PASS1 (dashed thin red line), PASS (black dashed
line) and HOLD integrator (dotted black lines). }
\label{fig_N100}
\end{figure}

\begin{figure}
\centering
\epsfig{file=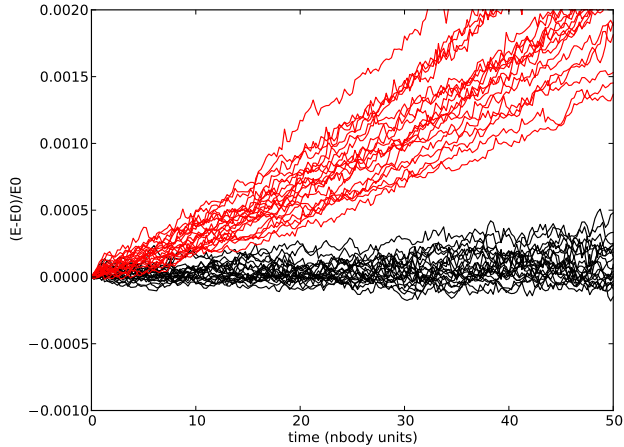, width=.7\textwidth}
\caption{Growth of relative energy error for 20 symmetrized (black lines) and 
20 non-symmetrized time step (red lines) runs. This is the same test as in
Makino et al. (2006). The same set of $N=100$ Plummer sphere initial conditions
were used for the two sets of simulations. Symmetrized runs took 
$\approx 20 \%$ longer than unsymmetrized runs (26 vs 22 sec in total).}
\label{fig_mak}
\end{figure}

To examine the differences between the 5 integrators presented in section
\ref{sec_der} we evolve an $N=100$ Plummer sphere of particles with equal
mass. Total mass is 1 and units are scaled such that G=1, total mass M=1 and
E=-0.25 \citep[Nbody units,][]{Heggie1986}.  Figure ~\ref{fig_N100} shows
the evolution of the Energy E, momentum P, angular momentum L and the center
of mass position X for each integrator, all evolved with the same
$\eta=0.01$. 

The first thing to notice in these plots is the different behavior of the
BLOCK integrator with respect to momentum conservation. The error in center of mass
and momenta (linear and angular) remain at machine precision for  
the conservative integrators, whereas the BLOCK integrator shows an error of
$\approx 10^{-7}$.  Note also that the center of mass of the PASS1 
integrator run is not conserved - this may be suprising: this integrator 
does conserve linear momentum because of the pairwise force kicks 
(as can be seen in fig.~\ref{fig_N100}). The reason for the shift in the center of 
mass is the fact that this integrator includes kicks between particles 
that are not synchronized in time: combining Eq.~\ref{eq_SF2nd} with 
Eq. \ref{eq_pass1} we see that midway in the half step of the evolution of $\exp{dt/2 \mathbb{H}_F}$ the 
kicks between the S and F system are calculated. At that point the positions 
of the S system particles lag $dt/4$ with respect to the F system. After the kicks 
are applied the position of an S particles drifts an extra $\Delta v_1 dt/4$  
from the first $\Delta v_1$ kick. In the second half step this is compensated by 
missing a $\Delta v_2 \Delta t/4$ drift, but these two will in general 
not cancel out exactly: hence the center of mass position is not conserved. 
This will happen irrespective of whether a DKD or 
KDK split is taken as the basis for PASS1. The proper formulation (PASS), which ensures
time synchronized kicks, does show conservation of the center of mass position 
(fig.~\ref{fig_N100}). Note also that applying the KDK method with the HOLD 
integrator  (for $\exp{dt \mathbb{H}_S}$ in  eq.~\ref{eq_hold}) will result 
in kicks between particles at different evolve times, and thus a deviation 
of the center of mass.  

The energy errors  are very similar for all the integrators. They remain 
all at a level of few$\times 10^{-7}$ and do not show any linear drift.  The
reason that the BLOCK integrator performs competitive with the other 
integrators is that the errors in momentum etc. remain small enough not to
induce energy errors (Note that the Plummer sphere initial conditions
represent a problem that is not especially sensitive to errors in momentum
conservation).

As we mentioned above, our symmetrization scheme (eq.~\ref{eq_symdt2}) 
of the time step is not exact, approximating the formal symmetry of 
Eq.~\ref{eq_symdt}. The Kepler test of the symmetrization scheme  
in section~\ref{sec_symtime} suggests that it is adequate, however 
for larger N simulations the additional terms in the timestep derivatives
in Eq.\ref{eq_symdt2} or the selection of minimum timesteps in Eq.~\ref{eq_symts}
could cause the timestep symmetrization to fail.
Hence, we conduct an additional test of our scheme by repeating the test of 
\cite{Makino2006} presented in their figure 4. This test consists of running 
$N=100$ Plummer sphere models for symmetrized and non-symmetrized 
time steps. \cite{Makino2006} showed that the linear drifts of the non-symmetrized
time step runs are suppressed when using the symmetrized version (they used
a scheme for block time steps using 6 iterations to symmetrize the
time step). Our result for this test is shown in figure~\ref{fig_mak}. The
integrator we use for both sets of runs is the HOLD integrator. The result
is very similar to the \cite{Makino2006} figure. This
shows that our symmetrization scheme works at least as well as the approach
given there,  and seems sufficient for the second order integrators 
employed here.

\begin{figure}
\centering
\epsfig{file=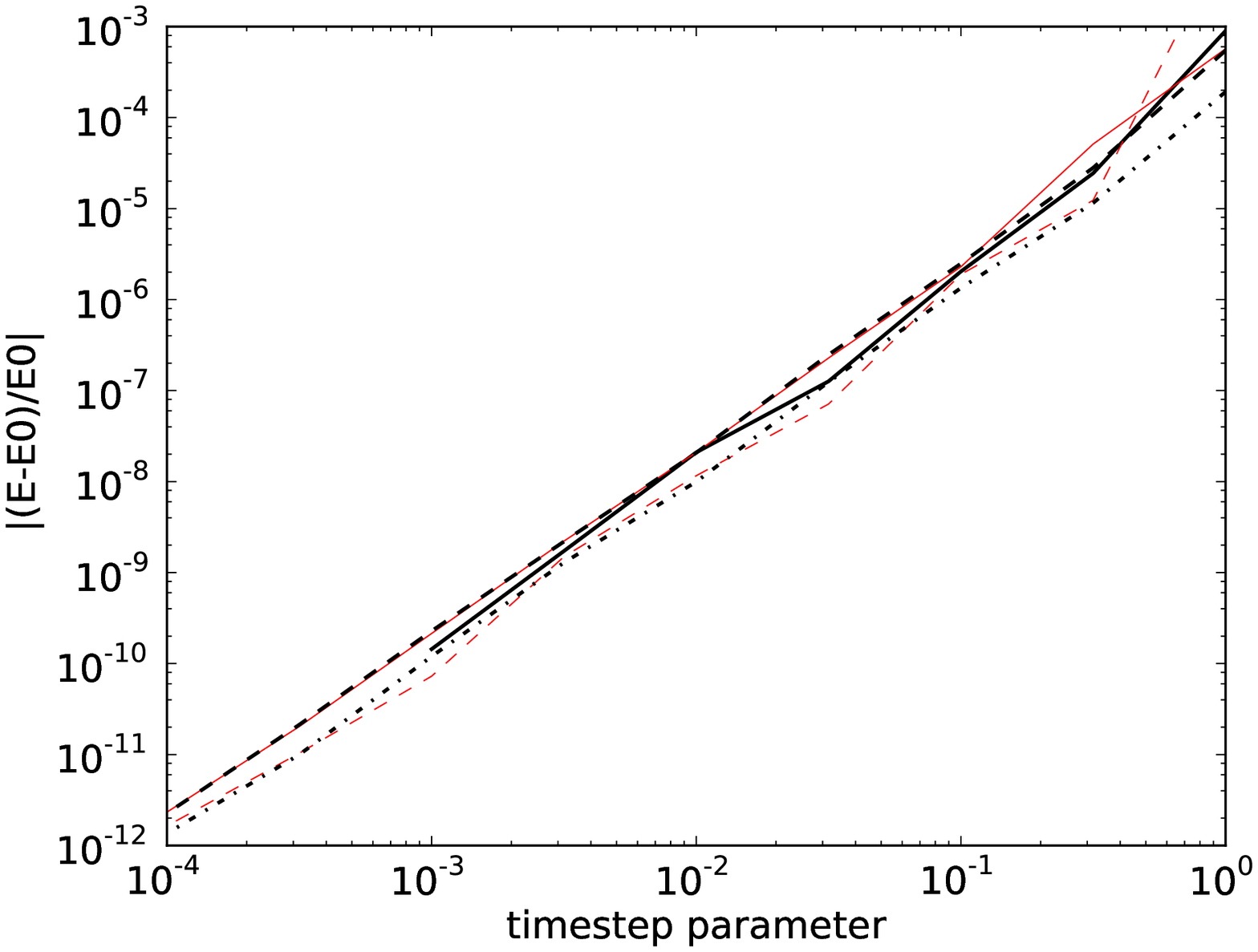, width=.49\textwidth}
\epsfig{file=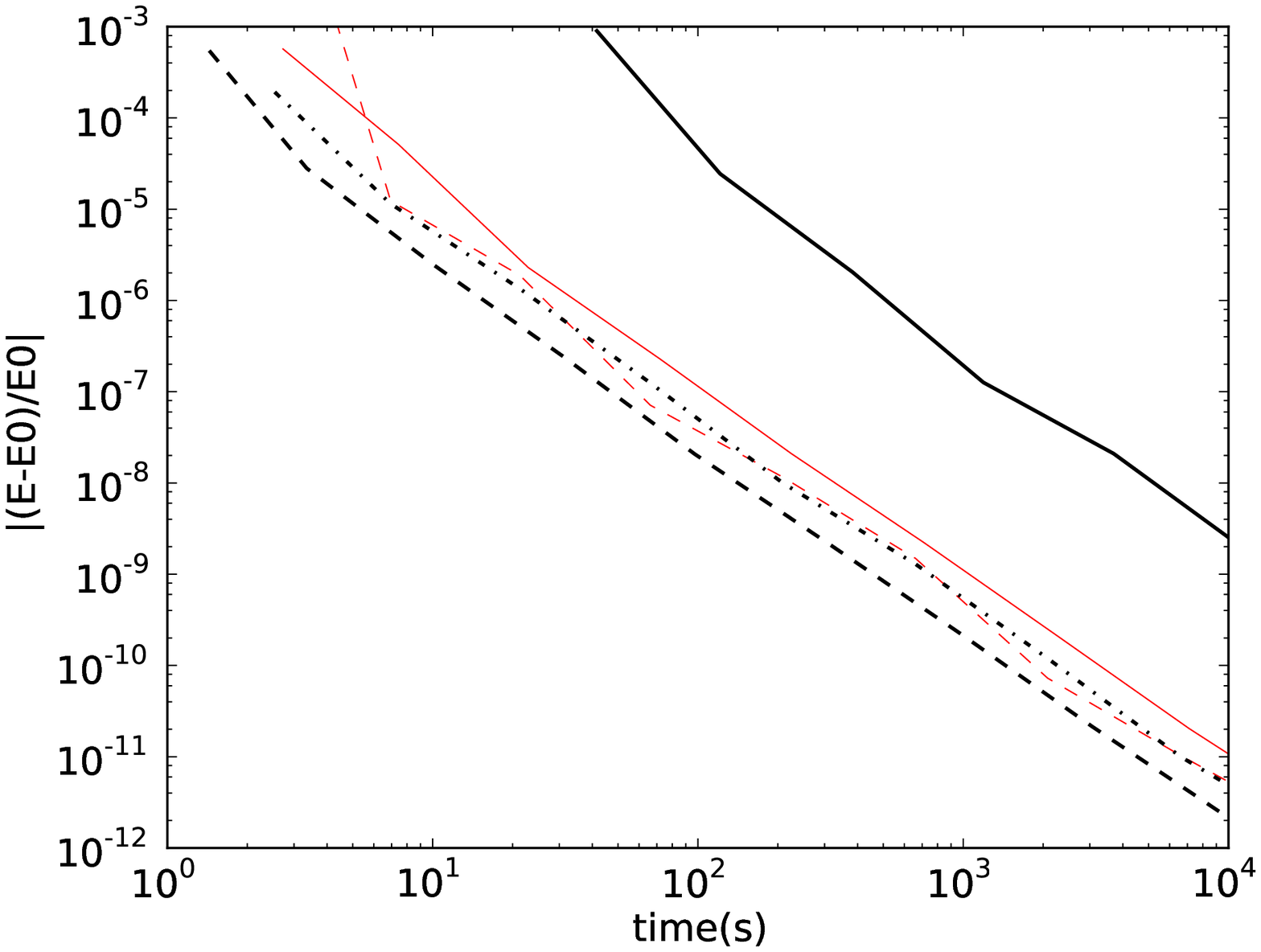, width=.49\textwidth}
\caption{Energy error as versus time step parameter and run time. Each run
evolves an $N=1024$ plummer sphere for 1 N-body time unit (lines use the
same legend as in fig~\ref{fig_N100}). Left panel shows relative energy error as
a function of time step parameter $\eta$, right panel shows error versus 
the wallclock time elapsed for that run. Energy error is
the maximum recorded over the run.}
\label{fig_eta}
\end{figure}

In figure~\ref{fig_eta} we show the behavior of the integrators as a
function of the time step parameter $\eta$. All integrators show the second
order dependence on $\eta$ as expected. The actual energy errors at the same
$\eta$ also lie within a narrow band. If we look at the trend with run-time 
we see that the SHARED integrator is a factor 300-1000 slower than the other
integrators. Within the integrators with individual time steps the HOLD
integrator is most efficient, then the PASS1 and PASS integrators and
lastly the BLOCK scheme. Looking at energy error dependence on $\eta$ we see
that BLOCK and HOLD are actually very similar, so the difference (of a
factor $\approx 3$) in efficiency as measured by wall-clock time 
is a result of the lower number of force evaluations of the HOLD method.

\subsection{Core collapse}

\begin{figure}
\centering
\epsfig{file=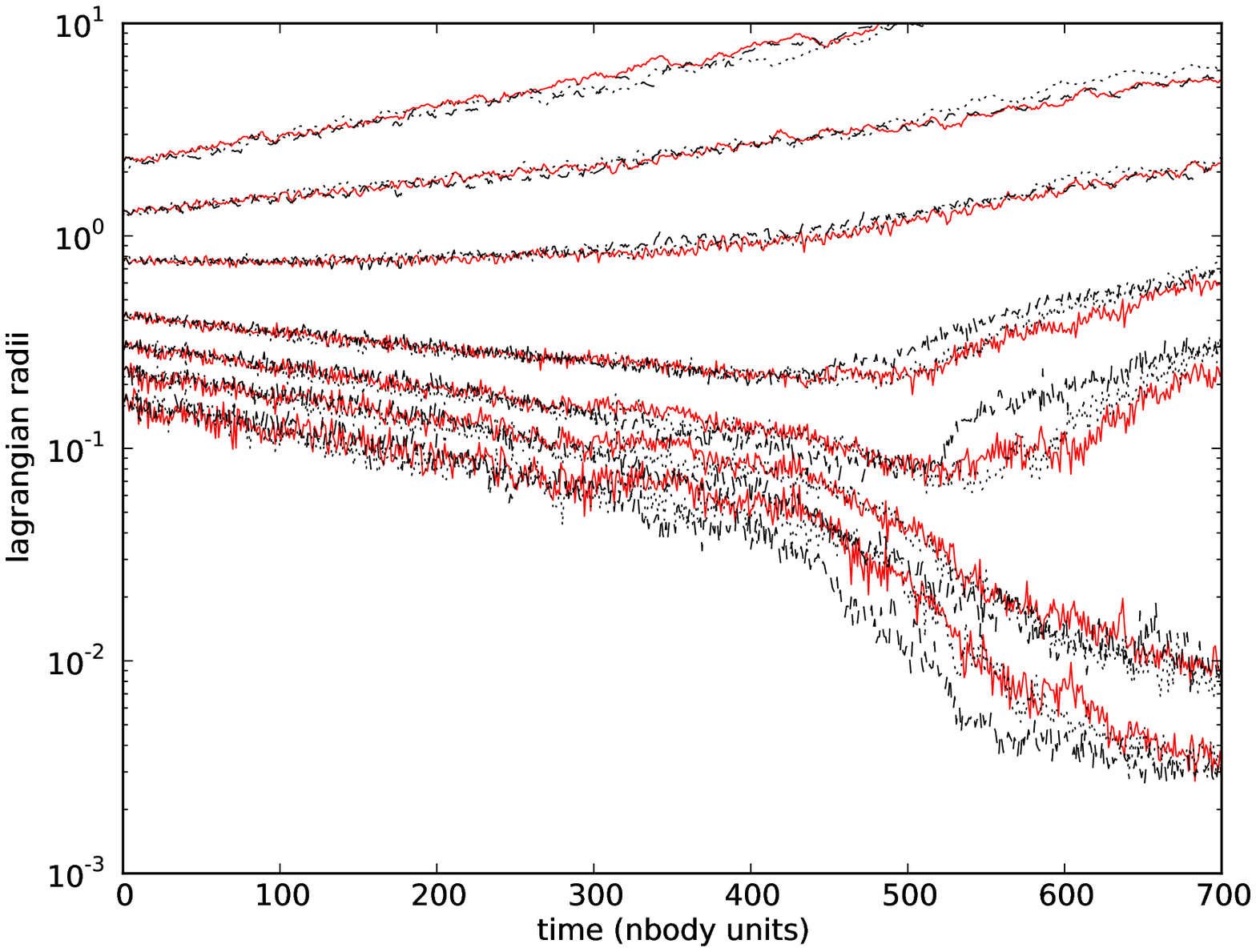, width=.49\textwidth}
\epsfig{file=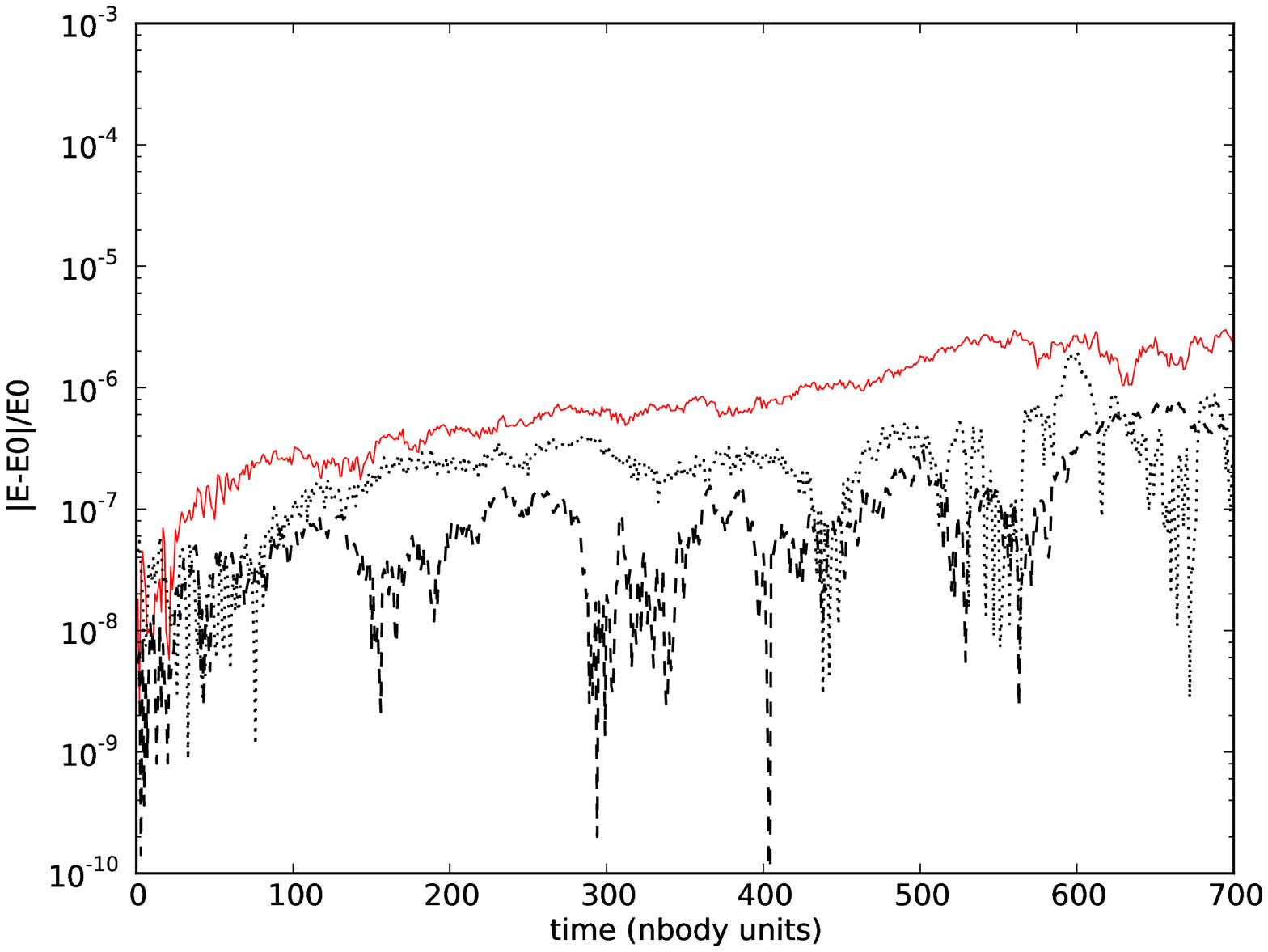, width=.49\textwidth}
\epsfig{file=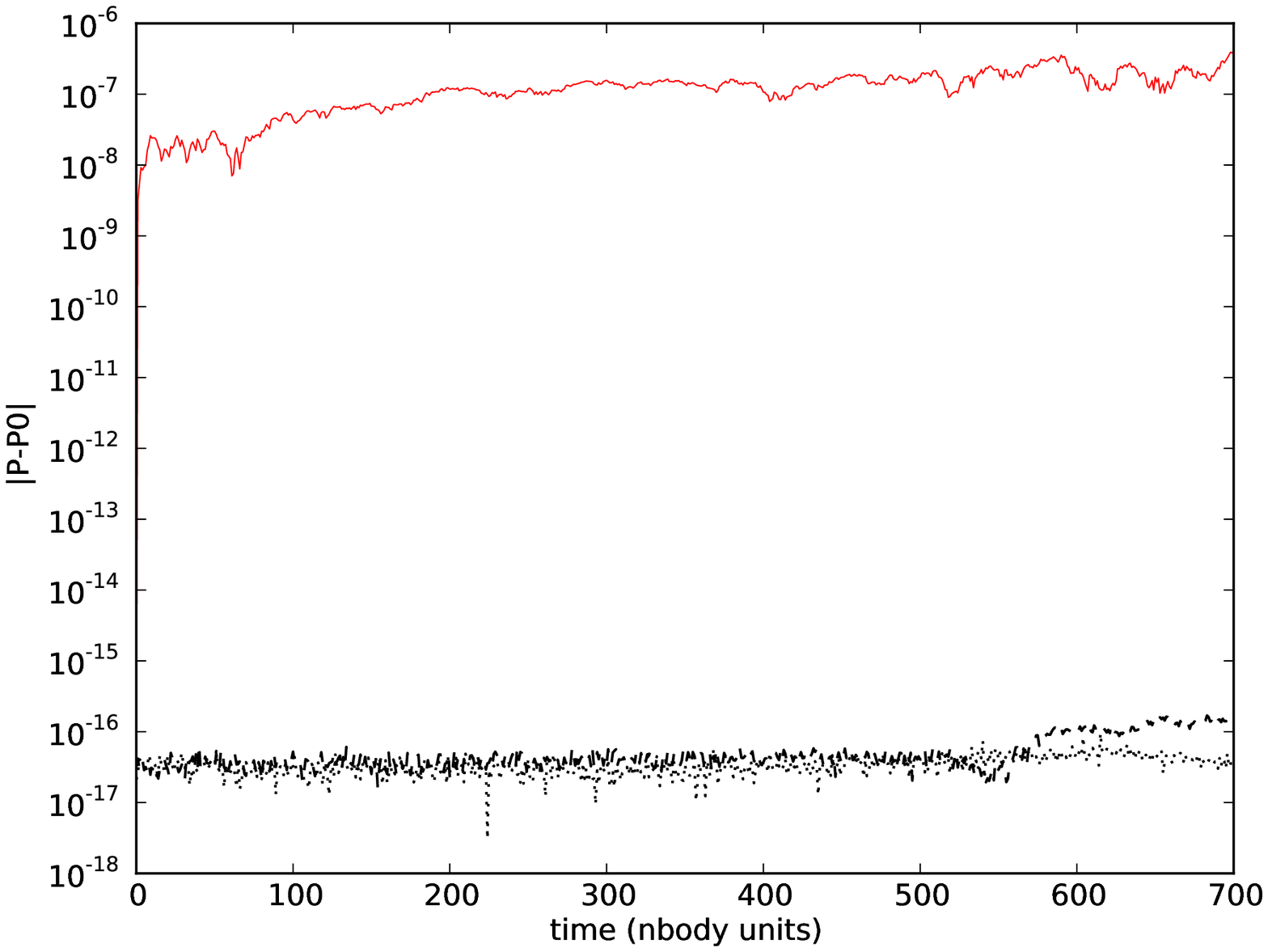, width=.49\textwidth}
\epsfig{file=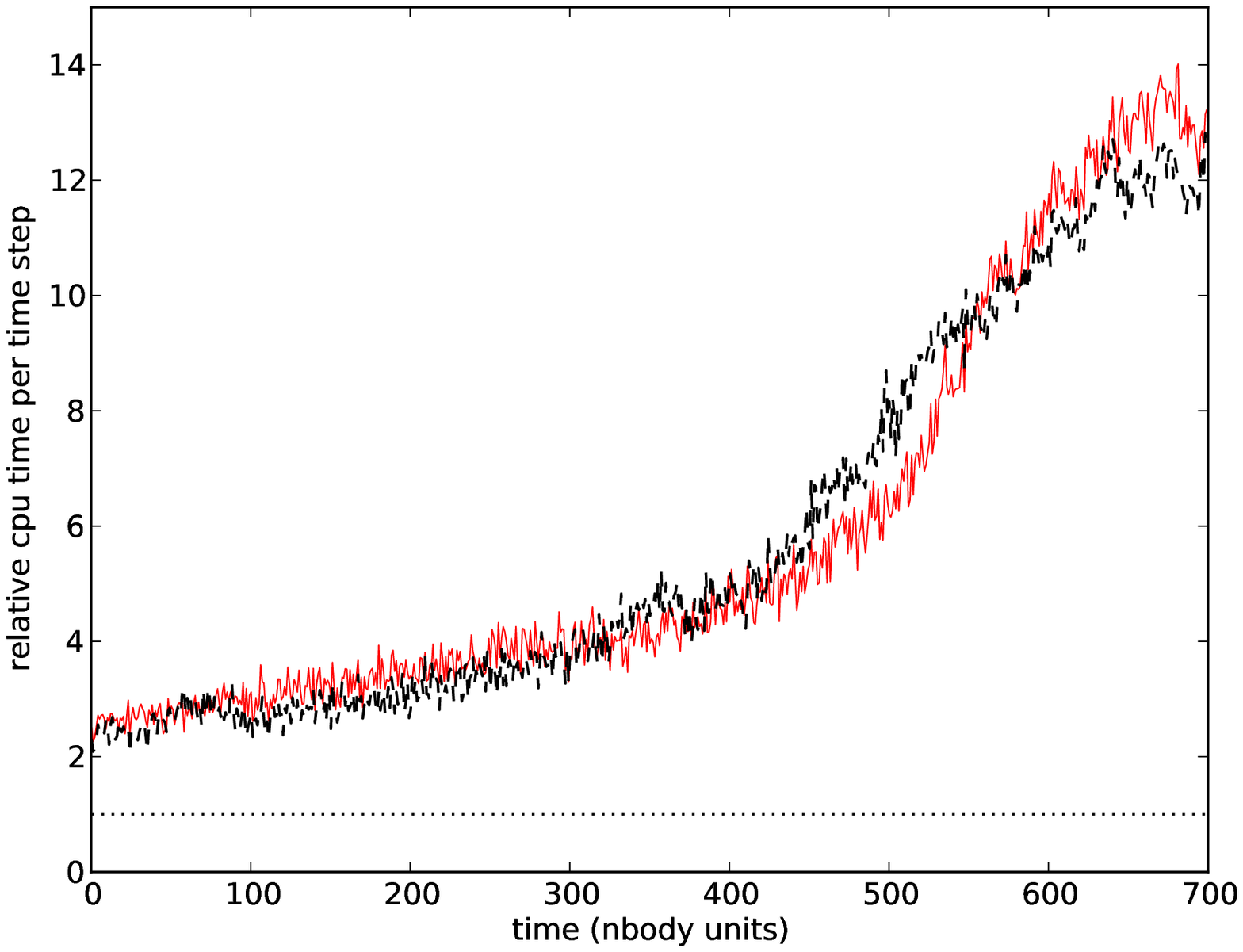, width=.49\textwidth}
\caption{
Lagrangian radii, energy and momentum momentum error and relative 
run-time for the N=1024 Plummer sphere core collapse test with smoothing. Upper left panel shows
Lagrangian radii for runs with the HOLD (dotted black lines), PASS  (dashed
black lines) and the BLOCK (drawn red lines) integrators (lines in  the
other panels use the same coloring).  Upper right panel shows the relative
energy error, lower left panel  shows the momentum error and lower right
shows the relative cpu time  per unit step for the simulation with respect
to the HOLD run. 
}
\label{fig_ccsmooth}
\end{figure}

\begin{figure}
\centering
\epsfig{file=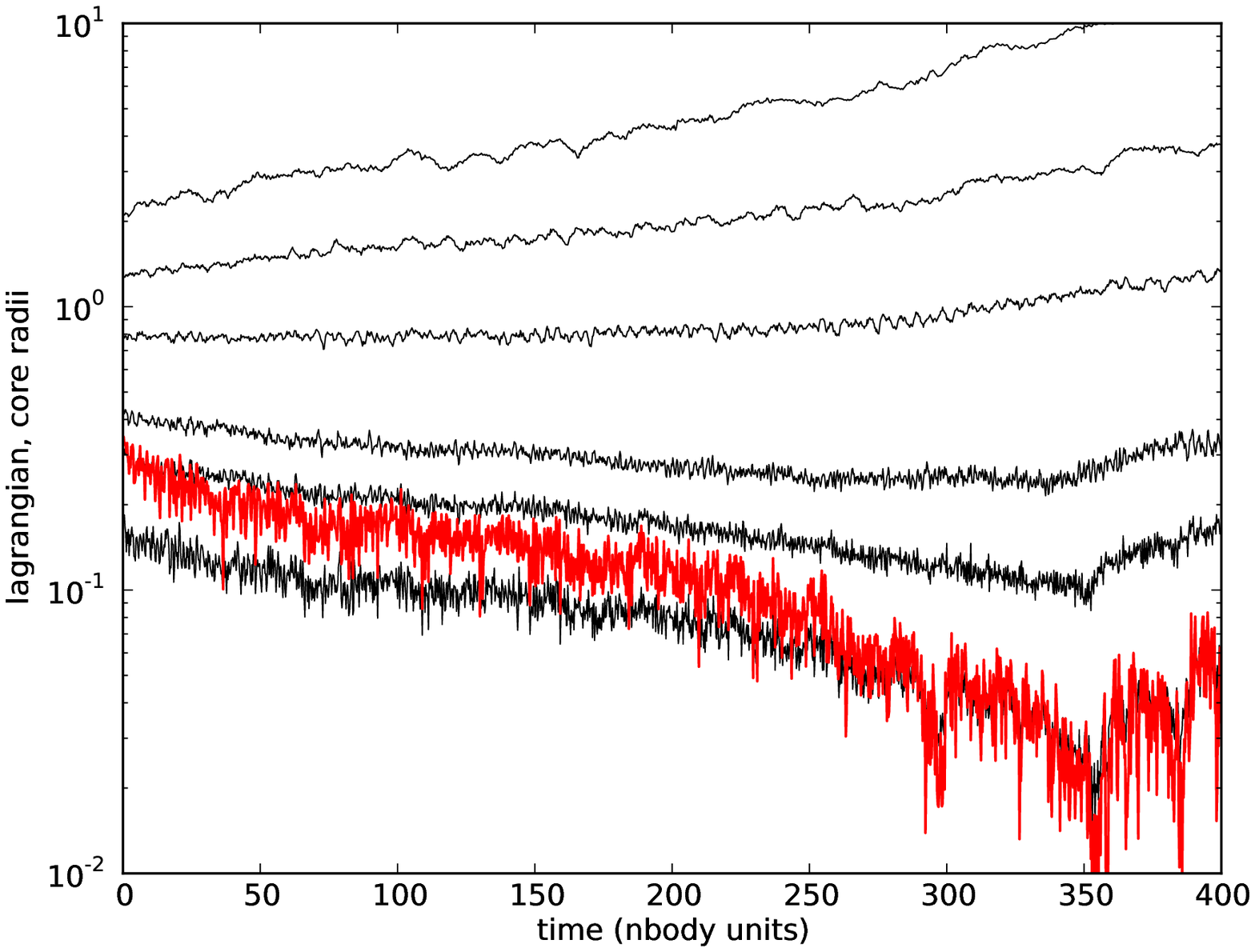, width=.55\textwidth}
\epsfig{file=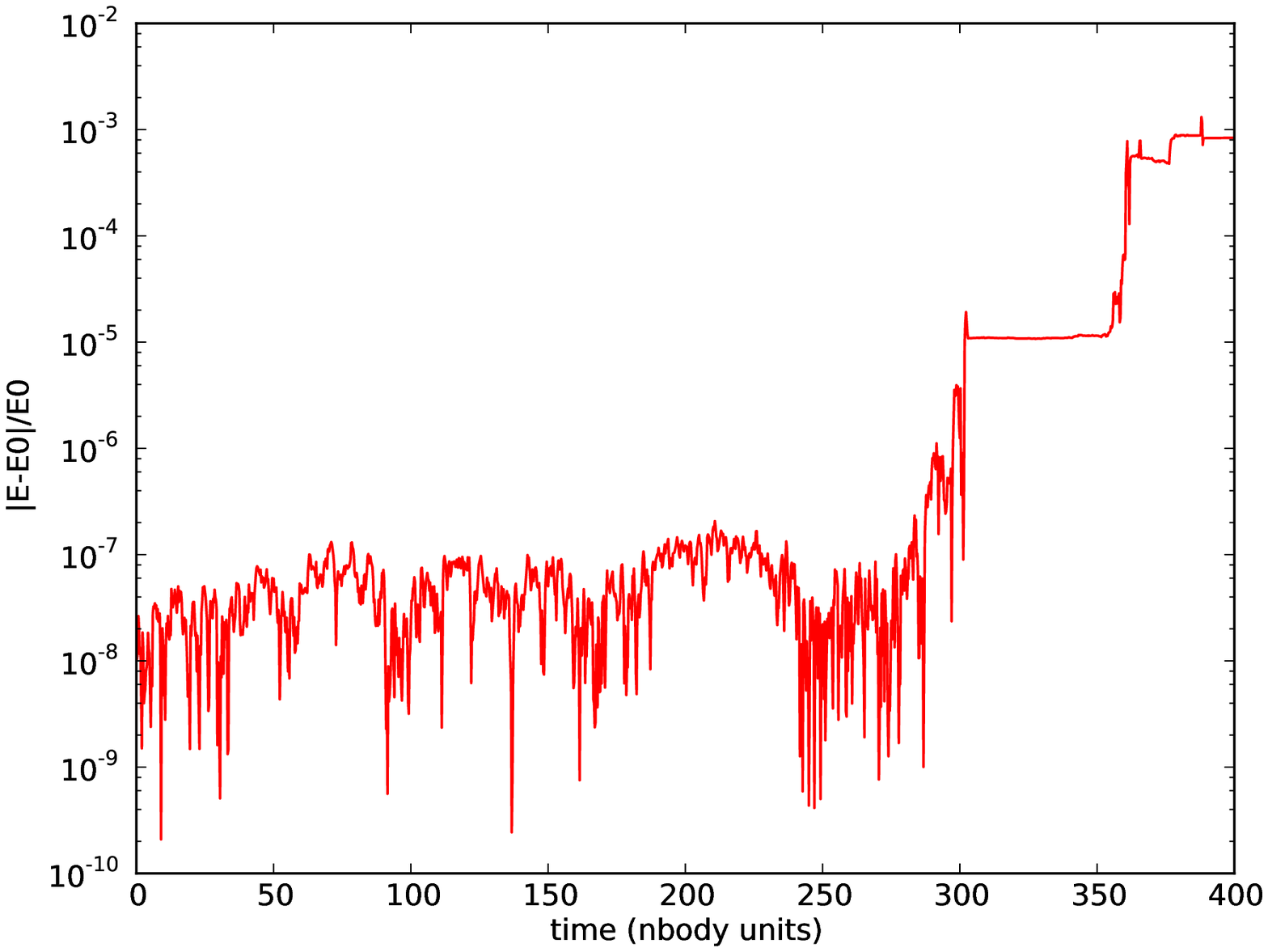, width=.55\textwidth}
\epsfig{file=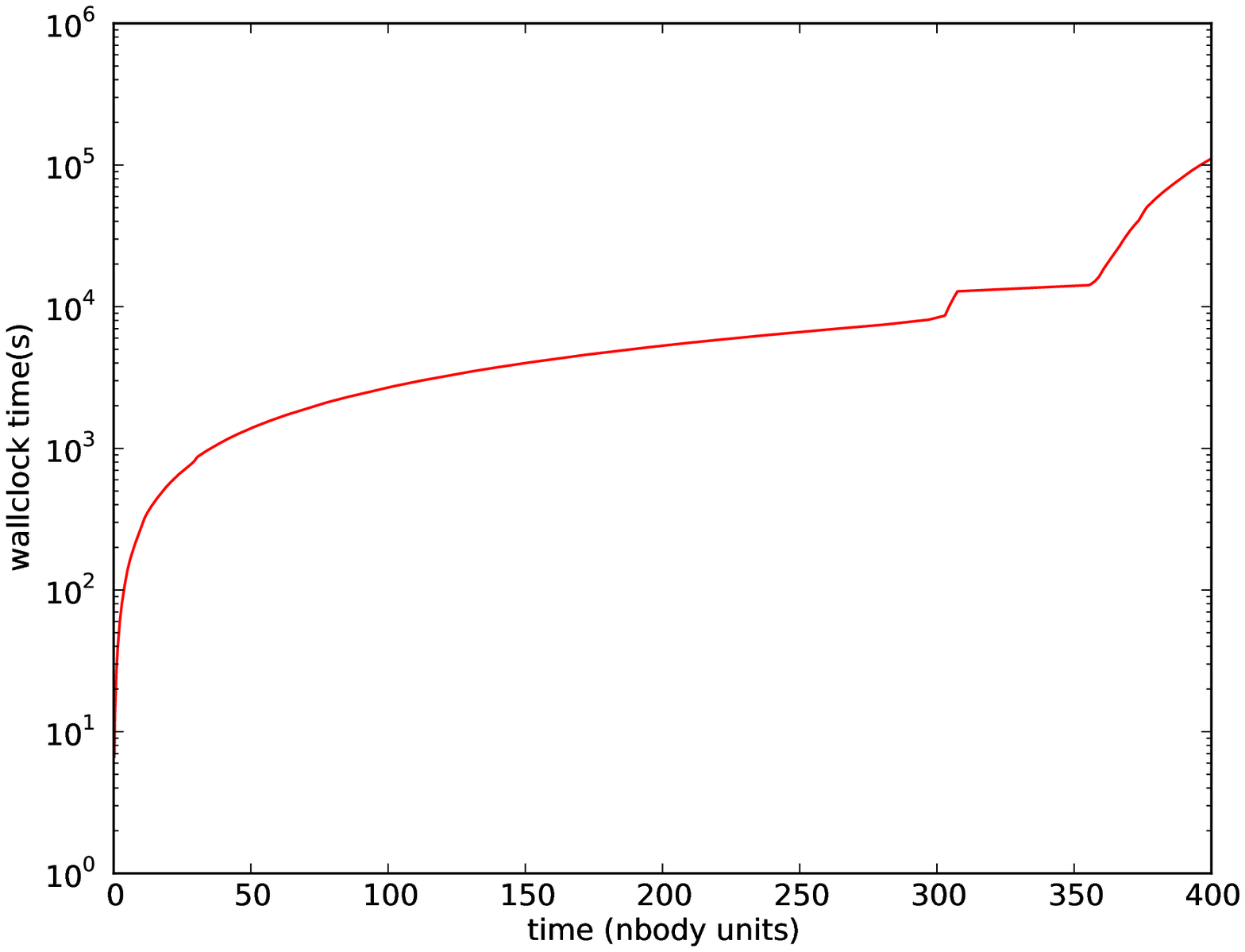, width=.55\textwidth}
\caption{Lagrangian radii, energy error and wallclock time for the N=1024 
Plummer sphere core collapse test with $\epsilon=0$ for the HOLD integrator. 
Upper panel shows lagrangian radii, Middle panel shows the relative energy 
error, lower panel shows the wall-clock time for the simulation. }
\label{fig_cc}
\end{figure}

As a last test we evolve a $N=1024$ Plummer sphere of equal mass particles
to core collapse. First we employ smoothing with an $\epsilon=1/256$
($\eta=0.01$). In this case we evolve to $t=700$. The setup was chosen to
match the test in \cite{Nitadori2008}. We plot the Lagrangian radii, energy
and momentum error in figure~\ref{fig_ccsmooth} (The angular momentum and
center of mass errors are omitted as they show the same pattern as the
momentum error). We plot the results for the PASS, BLOCK and HOLD
integrators.  As can be seen in the plot of the Lagrangian radii all
integrators show the expected evolution of the mass distribution.
Differences within the runs fall within the expected statistical variation
\citep[see also][]{Nitadori2008}. Note that although the runs start with the
exact  same initial conditions these can be considered effectively
different  runs as they diverge exponentially on the crossing time
timescale. The energy errors in Figure~\ref{fig_ccsmooth} show similar
results for the integrators, the BLOCK integrator showing a consistently
factor 2 to 3 larger error. Note the absence of energy drift also in this
long term integration (the energy errors do increase somewhat as the
integration progresses, but they return to zero and change sign).  In the
plot of the momentum error we can see that the PASS and HOLD integrators
conserve momentum to machine precision. A slight increase in error is
visible at the end. This is due to the loss of precision in calculations
involving escaping stars. If we compare the time needed to advance the
simulation of the three integrators we notice a clear difference: at the
beginning the HOLD integrator is a factor $\approx 3$ faster (which is
consistent with the tests in fig.~\ref{fig_N100}), as the simulation evolves
the difference between HOLD  and the other two integrators increases to a
factor 12 at the end. The HOLD integrator becomes more efficient as the
timescale differences within the system become larger.  

As an additional test for the HOLD integrator we run the same problem for
$\epsilon=0$. This is a demanding test as in this case strongly bound
binaries  form during core collapse \citep{Giersz1994}. This test
calculation is shown in figure~\ref{fig_cc}. In this case the core collapse
occurs at $t\approx 350$, and the evolution of the Lagrangian radii and core
radius shows a pattern consistent with \cite{Giersz1994}. The energy errors
increase during the core collapse  \citep[for reference this can be compared
with current Hermite codes in][their fig. 21]{Konstantinidis2010}. Note that  
the error increases during very short
episodes, after which the error remains bounded. These error jumps occur during 
strong interactions with the binary that forms in the center, and may be caused 
by failures of the timestep symmetrization, although similar error jumps occur also in codes
using completely different timestepping criteria \citep[see again][for an example]{Konstantinidis2010}
A more uniform error behavior may be obtained by decreasing the $\eta$ parameter
during these 'rough patches,'  for a modest increase in computation time 
(for very close encounters the loss of accuracy
may be due to the numerical precision, in this case some form of regularization
is necessary).
Plotting the wall-clock time of the simulation we see that towards the end of
the simulations slows down (for the BLOCK and PASS integrators this would
mean an additional slowdown by a factor 100). This is due to the formation
of hard binaries. The HOLD integrator therefore is able to integrate systems
without smoothing,  although a more efficient handling of binaries is
desirable. 

\section{Discussion}
\label{sec_disc}

We have presented a set of conservative individual time step integrators 
derived using recursive Hamiltonian splitting. These integrators are
remarkably simple and closely follow  previous work: specifically they can
be seen as a generalization of the scheme used in the mesh-based integrator
for collision-less systems  GROMMET~\citep{Magorrian2007} and are closely
related to the Hamiltonian splitting methods used in planetary
dynamics \citep{Duncan1998, Saha1994} or the force splitting methods used in molecular
dynamics \citep[e.g.][]{tuckerman1992}. In contrast  to the usual formulation of
block time steps they conserve momentum and  angular momentum to machine
precision and in conjunction with the (approximately) symmetrized timstep criterion presented
here they show good conservation of energy. Our tests show that the  energy
conservation is a consequence of the use of a symmetrized time step 
criterion (as the conventional BLOCK integrator shows similar error behavior 
with the symmetrized time step). Of the new integrators HOLD and PASS the 
former is more efficient because it calculates the interaction between two 
particles with different time steps at an interval determined by the 
slower of the two particles. The Hamiltonian splitting considered here 
can be applied more generally. For example, Smooth Particle Hydrodynamics
(SPH) codes also use a similar time-stepping scheme, and the scheme
presented here directly carries over to SPH codes. 

One drawback of the subdivision in terms of the time steps assigned to the
particles is that the interactions for a given particle in a time
bin are calculated together with all other particles with the same time step.
For example, if the system contains multiple binary stars (and assume all
these binaries have the same orbital parameters) then the stars that form
these binaries all have the same time step and will end up in the same bin,
with all their mutual interactions calculated on the fastest time step. The
worst case scenario (which is however physically significant) is that all the
stars are in binaries. In this case the split time step scheme as presented
here does not save computing time with respect to a shared time step
simulation (this is also the case for conventional block time step
integrators). To ameliorate this one would need a subdivision at each level
of the system in multiple systems based on e.g. a spatial partitioning at
each level, as opposed to a simple split in a fast and a slow system. 

We have limited ourselves to second order integrators. The second order
nature of the integrators is a consequence of Eq.~\ref{eq_SF2nd}: this is
the second order leapfrog composition. There is no reason why we could not 
use a higher order composition scheme \citep{Mclachlan1995}, as long as we take
care selecting the same (or at least of the same order) composition schemes
for the component integrators. A problem with using higher order integrators
is that the compositions contain terms with negative time steps. While in
principle this is not a problem, it does present a problem for the
efficiency of the integrator as the time interval that the integrator
follows is bigger than the integration time interval by a factor $\alpha>1$.
Note that within a recursive scheme every level in the time step hierarchy will   
increase the length of this interval by this factor, so that unless $\alpha$
is very close to 1, $\alpha^k $ can increase rapidly.

In addition to the algorithmic enhancements mentioned above, the numerical
implementation of the integrators can be improved. For large scale
production work they would need to be parallelized and probably also be
adapted to use Graphic Processor Units (GPUs).  At the moment a partial
implementation on GPUs is available, where only the kicks are implemented on
the GPU. This implementation suffers from the inefficiencies of transferring
data to and from the GPU. With these improvements we expect that the code
will be competitive with modern Hermite codes. 
\ \\ \\
{\bf Acknowledgements}
This work was supported by the Netherlands Research Council NWO 
(grants \#643.200.503, \#639.073.803 and \#614.061.608) and by 
the Netherlands Research School for Astronomy (NOVA).

\bibliographystyle{astron}
\bibliography{ms}

\end{document}